\documentclass[a4paper]{article}
\usepackage{amsmath}
\usepackage{amsfonts}
\usepackage{amssymb}
\usepackage{graphicx}
\usepackage{verbatim}
\usepackage[hidelinks]{hyperref}

\usepackage{geometry}
\usepackage{authblk}
\title{Cross-Cutting Political Awareness through \\Diverse News Recommendations \footnote{To be presented at \emph{European Symposium Series on Societal Challenges
in Computational Social Science 2019} in Zurich, Switzerland, September 2nd-4th, 2019.}}

\author[1]{Bibek Paudel\footnote{Work done partially while at University of Zurich, Zurich, Switzerland.}}
\author[2]{Abraham Bernstein}
\affil[1]{Stanford University, California, USA}
\affil[2]{University of Zurich, Zurich, Switzerland}
\date{}                     
\begin{document}
  \maketitle
\pagenumbering{gobble}

\hspace{10pt}

\normalsize
\textbf{Introduction.}
The Internet is now the primary source of political news in many Western countries~\cite{mitchell2015millennials} including Switzerland~\cite{uzhmediares}, offering people an ever-growing pool of news sources.
It also influence the norms, opinions, actions, and engagements that constitute democratic engagement. 
However, the abundance of online news sources has been found to lead to information overload and feelings of inability to cope with the news flow in people of all ages, ultimately resulting in ``news fatigue''~\cite{nytfatigue}. 
To deal with this, users often rely on recommender systems and Online Social Networks (OSNs) to select news for consumption~\cite{aljukhadar2012using}.
Recommender systems find extensive use in social networks, media platforms and search engines, and influence user's exposure to news.
This makes the role of such technologies even more important, especially in major political events and discussions.
They also play a large role in shaping opinions and setting the agenda for more traditional news media. 


However, the suggestions generated by most existing recommender systems are known to suffer from a lack of diversity, and other issues like popularity bias~\cite{paudel2017fewer, paudel2017updatable}.
As a result, they have been observed to promote well-known ``blockbuster'' items, and to present users with ``more of the same'' choices that entrench their existing beliefs and biases~\cite{sunstein2001echo}.
This limits users' exposure to diverse viewpoints and potentially increases political polarization.
%
%
%
The underlying reason is that these algorithms promote items that match a user's previous choices or are popular among similar users.
They are designed to optimize prediction accuracy and click rates, measured in terms of how well they can identify items similar to the past observations. 

This has far-reaching impacts on the social and democratic processes, as mass media heavily influence 
opinion formation and the level of civic engagement of citizens as well as political actors.
For this reason, we see the need for better ways to recommend political news to readers.
Given the importance of high-quality as well as balanced news consumption for reducing political polarization and a functioning democracy~\cite{muller2014comparing,mansbridge2012systemic,coronel2003role}, it is important to investigate ways to improve the reader's ability to filter this continuous flow effectively and to provide them with diverse viewpoints for cross-cutting discussions~\cite{mutz2002cross}.

\textbf{Our approach for recommending politically diverse content.} To \emph{promote the diversity of views}, we developed a novel computational framework that can identify the political leanings of users and the news items they share on OSNs.
Based on such information, our system can recommend news items that \textit{purposefully expose users to different viewpoints and increase the diversity of their information ``diet.''}
Our research on recommendation diversity and political polarization helps us to develop algorithms that measure each user's reaction 
and adjust the recommendation accordingly. The result is \emph{an approach that exposes users to a variety of political views and will, hopefully, broaden their acceptance (not necessarily the agreement) of various opinions}.

A common way to diversify content is by including viewpoints from different sources, assuming that ideological positions of political elites and news providers are fixed over long durations.
In events like the Brexit referendum, this approach is likely to suffer from a major problem.
A set of viewpoints from politicians or news sources belonging to different parties with different ideologies 
could still be homogeneous if it includes viewpoints from the member of the two parties supporting the same campaign during events like the Brexit referendum~\cite{gamble2019realignment,whitefield2018issue}. The authors in~\cite{gamble2019realignment} notice that ``the Conservatives are a party of Brexiteers led by a Remainer, while Labour is a party of Remainers led by a Brexiteer.''
The left-right classification of users or politicians based on their long-term behavior like speeches, voting habits or social media follower patterns is also likely to fall short.

As a first step towards tackling this problem, our approach incorporates ideological positions about particular political events learned from social media signals, instead of relying on long-term or general political alignments. Our proposed solution has two components: (i) learning ideological positions of users, political elites, and web content in the context of specific political events, (ii) using the ideological positions to diversify recommendation based on a diversification strategy similar to our previous work~\cite{christoffel2015blockbusters}.

Previous research has shown that social media behavior can be used to predict the ideological positions of political actors~\cite{conover2011predicting,barbera2013}.
Our work goes further and uses this idea to generate diverse news recommendations.
Our approach incorporates multiple social media signals like the sharing of news stories to find the \emph{ideological positions} of not only users, but also of individual news stories, video, or other material shared on social media more accurately.
As a result we can categorize items and users in terms of their similarity of ideological position.
Based on these positions, our recommendation strategy then recommends political content that expose users to different viewpoints of varying ideological stance. This can be used to increase the diversity of their information diet. 

It is not sufficient to just expose users to diverse viewpoints~\cite{bail2018exposure}.
It is also important to identify the types of information users are already familiar with, and what new viewpoints will be agreeable to them.
Based on the insight that chance encounters via weak ties lead to diversity~\cite{prior2007post,sunstein2002republic}, we use the estimated ideological positions to recommend news-items that are not deemed completely disagreeable to users, but are more diverse than those recommended by existing systems. We use a modified random-walk based exploration of the user-item feedback graph in order to diversify recommendations through weak-ties~\cite{paudel2017fewer,paudel2017updatable}.

As a result, our system can nudge users to different viewpoints, and hopefully change the nature of information being consumed and spread on social networks. In other words: we hope to \textit{gradually widen a person's news exposure and reduce political polarization without driving them off with disagreeable content}.

\textbf{Current state of research and future work}.
We collected large datasets of Twitter discussions related to major political events in three Western countries: US Presidential Elections of 2016, Brexit referendum of 2016, and German Federal Elections of 2017.
With experimental evaluations on these social network datasets, we find that our method is able to generate more politically diversified recommendations to the users without overly sacrificing prediction accuracy. A research paper based on this work is currently under submission.
As a next step, we are using these insights to deploy and test them in a large-scale experiment involving multiple news producers and consumers.

\bibliographystyle{abbrv}
{\footnotesize
\bibliography{bibliography}}

\begin{thebibliography}{10}

\bibitem{aljukhadar2012using}
M.~Aljukhadar, S.~Senecal, and C.-E. Daoust.
\newblock Using recommendation agents to cope with information overload.
\newblock {\em International Journal of Electronic Commerce}, 17(2):41--70,
  2012.

\bibitem{bail2018exposure}
C.~A. Bail, L.~P. Argyle, T.~W. Brown, J.~P. Bumpus, H.~Chen, M.~F. Hunzaker,
  J.~Lee, M.~Mann, F.~Merhout, and A.~Volfovsky.
\newblock Exposure to opposing views on social media can increase political
  polarization.
\newblock {\em Proceedings of the National Academy of Sciences},
  115(37):9216--9221, 2018.

\bibitem{barbera2013}
P.~Barber{\'a}.
\newblock Birds of the same feather tweet together. bayesian ideal point
  estimation using twitter data.
\newblock {\em Political Analysis}, 2013.

\bibitem{christoffel2015blockbusters}
F.~Christoffel, B.~Paudel, C.~Newell, and A.~Bernstein.
\newblock Blockbusters and wallflowers: Accurate, diverse, and scalable
  recommendations with random walks.
\newblock In {\em Proceedings of the 9th ACM Conference on Recommender
  Systems}, pages 163--170. ACM, 2015.

\bibitem{conover2011predicting}
M.~D. Conover, B.~Gon{\c{c}}alves, J.~Ratkiewicz, A.~Flammini, and F.~Menczer.
\newblock Predicting the political alignment of twitter users.
\newblock In {\em Privacy, Security, Risk and Trust (PASSAT) and 2011 IEEE
  Third Inernational Conference on Social Computing (SocialCom), 2011 IEEE
  Third International Conference on}, pages 192--199. IEEE, 2011.

\bibitem{coronel2003role}
S.~Coronel.
\newblock The role of the media in deepening democracy.
\newblock {\em NGO Media Outreach: Using the}, 2003.

\bibitem{gamble2019realignment}
A.~Gamble.
\newblock The realignment of british politics in the wake of brexit.
\newblock {\em The Political Quarterly}, 2019.

\bibitem{mansbridge2012systemic}
J.~Mansbridge, J.~Bohman, S.~Chambers, T.~Christiano, A.~Fung, J.~Parkinson,
  D.~F. Thompson, and M.~E. Warren.
\newblock A systemic approach to deliberative democracy.
\newblock {\em Deliberative systems: Deliberative democracy at the large
  scale}, pages 1--26, 2012.

\bibitem{mitchell2015millennials}
A.~Mitchell, J.~Gottfried, and K.~E. Matsa.
\newblock Millennials and political news.
\newblock {\em Pew Research Center}, 1, 2015.

\bibitem{muller2014comparing}
L.~M{\"u}ller.
\newblock {\em Comparing mass media in established democracies: patterns of
  media performance}.
\newblock Springer, 2014.

\bibitem{mutz2002cross}
D.~C. Mutz.
\newblock Cross-cutting social networks: Testing democratic theory in practice.
\newblock {\em American Political Science Review}, 96(1):111--126, 2002.

\bibitem{nytfatigue}
{New York Times}.
\newblock Fatigued by the news? {E}xperts suggest how to adjust your media
  diet.
\newblock
  \url{https://www.nytimes.com/2017/02/01/us/news-media-social-media-information-overload.html},
  2017.
\newblock Accessed: 2017-12-20.

\bibitem{paudel2017updatable}
B.~Paudel, F.~Christoffel, C.~Newell, and A.~Bernstein.
\newblock Updatable, accurate, diverse, and scalable recommendations for
  interactive applications.
\newblock {\em ACM Transactions on Interactive Intelligent Systems (TiiS)},
  7(1):1, 2017.

\bibitem{paudel2017fewer}
B.~Paudel, T.~Haas, and A.~Bernstein.
\newblock Fewer flops at the top: Accuracy, diversity, and regularization in
  two-class collaborative filtering.
\newblock In {\em Proceedings of the Eleventh ACM Conference on Recommender
  Systems}, pages 215--223. ACM, 2017.

\bibitem{prior2007post}
M.~Prior.
\newblock {\em Post-broadcast democracy: How media choice increases inequality
  in political involvement and polarizes elections}.
\newblock Cambridge University Press, 2007.

\bibitem{sunstein2001echo}
C.~R. Sunstein.
\newblock {\em Echo chambers: Bush v. Gore, impeachment, and beyond}.
\newblock Princeton University Press Princeton, NJ, 2001.

\bibitem{sunstein2002republic}
C.~R. Sunstein.
\newblock {\em Republic. com}.
\newblock Princeton University Press, 2002.

\bibitem{uzhmediares}
{University of Zurich}.
\newblock Internet is the primary source of information in switzerland.
\newblock
  \url{http://www.media.uzh.ch/en/Press-Releases/archive/2015/schweizer-informieren-sich-hauptsaechlich-ueber-das-internet.html},
  2015.
\newblock Accessed: 2017-12-20.

\bibitem{whitefield2018issue}
S.~Whitefield and J.~Rovny.
\newblock Issue dimensionality and party politics in turbulent times.
\newblock {\em Party Politics}, 2018.

\end{thebibliography}

\end{document}